\documentclass[12pt]{iopart}

\usepackage{graphics}
\usepackage{graphicx}
\usepackage{amssymb}
\usepackage{epsfig}
\usepackage{color}
\usepackage{ulem}
\usepackage{arydshln}
\usepackage{multirow}

\begin{document}
\title[Nature of the magnetic ground state of CeRuSn]{Nature of the magnetic ground state in the mixed valence compound CeRuSn: a single-crystal study}

\author{J Fik\'{a}\v{c}ek$^1$, J Prokle\v{s}ka$^1$, J Prchal$^1$, J. Custers$^1$, V Sechovsk\'{y}$^1$}

\address{$^1$ Faculty of Mathematics and Physics, Charles University in Prague, Ke Karlovu 5, 121 16 Prague 2, Czech Republic.}
\ead{janfikacek@gmail.com}

\begin{abstract}
We report on detailed low temperature measurements of the magnetization, the specific heat and the electrical resistivity on high quality CeRuSn single crystals. The compound orders antiferromagnetically at $T_{\rm N} = 2.8$~K with the Ce$^{3+}$ ions locked within the $a-c$ plane of the monoclinic structure. Magnetization shows that below $T_{\rm N}$ CeRuSn undergoes a metamagnetic transition when applying a magnetic field of 1.5 and 0.8~T along the $a$ and $c$--axis, respectively. This transition manifests in a tremendous negative jump of $\sim 25$~\% in the magnetoresistance. The value of the saturated magnetization along the easy magnetization direction ($c$--axis) and the magnetic entropy above $T_{\rm N}$ derived from specific heat data correspond to the scenario where only one third of the Ce ions in the compound being trivalent and carrying a stable Ce$^{3+}$ magnetic moment, whereas the other two thirds of the Ce ions are in a nonmagnetic tetravalent and/or mixed valence state. This is consistent with the low temperature CeRuSn crystal structure i.\,e.\,, a superstructure consisting of three unit cells of the CeCoAl-type piled up along the $c$--axis, and in which the Ce$^{3+}$ ions are characterized by large distances from the Ru ligands while the Ce-Ru distances of the other Ce ions are much shorter causing a strong 4{\it f\/}-ligand hybridization and hence leading to tetravalent and/or mixed valence Ce ions.
\end{abstract}
\submitto{\JPCM}
\maketitle

\section{Introduction}

Although the 4{\it f\/} electron resides within the cerium atom core, its spatially extended 4f-electron wave function can hybridize with valence states of ligands in a compound. Depending on the 4{\it f\/}--ligand hybridization~\cite{1,2,3} strength, the trivalent state, which is standard for a majority of lanthanides, can be dynamically mixed with the itinerant 4{\it f\/}$^{4+}$ state to a various extent. The 4{\it f\/}--ligand hybridization strength is intimately connected with the overlap of the Ce 4{\it f\/}--electron wave function with the ligand valence electron wave functions and thus with the distance between the Ce ion and its ligands~\cite{2,3} (we label it by $d_{\rm{Ce-ligand}}$). When $d_{\rm{Ce-ligand}}$ is sufficiently large, the 4{\it f\/}--ligand hybridization is negligible and the Ce$^{3+}$ valence is preserved. The Ce$^{3+}$ ion bears a magnetic moment intrinsic to one localized 4{\it f\/} electron.  With decreasing $d_{\rm{Ce-ligand}}$, the 4{\it f\/} and ligand valence electron wave-functions overlap more and the enhanced hybridization results in a certain probability for the Ce ion to lose its only 4{\it f\/} electron and to become tetravalent. The admixture of probabilities of  Ce$^{3+}$ and Ce$^{4+}$ states is known as the intermediate valence state (also fluctuating valence state -- frequently labeled as 4{\it f\/}$^{4-\delta}$). The Ce intermediate valence ions typically do not carry any stable magnetic moment and thus do not participate in magnetic ordering~\cite{4,5,6}.\\
The Ce$^{3+}$ magnetic moments in metals are correlated and consequently the low temperature magnetic properties are governed either by direct or by the indirect exchange interaction, namely the ubiquitous RKKY interaction~\cite{7,8}, which is mediated by conduction electrons. Sufficiently strong exchange interactions yield a magnetically ordered ground state~\cite{9}. Contrary, these interactions compete with the on-site Kondo interaction. Here, conduction electrons form a singlet state with the Ce$^{3+}$ ion resulting in an effective screening of the Ce$^{3+}$ magnetic moment. Consequently a nonmagnetic ground state is established~\cite{9}.\\
CeRuSn has several polymorphs. At room temperature it adopts a superstructure of the monoclinic CeCoAl-type with two crystallographically independent cerium sites differing considerably in the distance to the Ru ligands ($d_{\rm{Ce-Ru}}$). The Ce site with long $d_{\rm{Ce-Ru}}$ can be assigned to a Ce$^{3+}$ ion, whereas the other site with short $d_{\rm{Ce-Ru}}$ was suggested to be occupied by an intermediate valent or tetravalent cerium~\cite{10,11}. This situation may give rise to a 4{\it f\/}--charge density wave with the periodicity of the superstructure~\cite{10,11,12}.
When cooling below room temperature, CeRuSn undergoes two successive polymorphic transitions to other superstructures of the CeCoAl-type structure characterized by various combinations of Ce-Ru ($d_{\rm{Ce-Ru}}$) distances along the $c$--axis, while the relative occurrence of the short Ce-Ru distances increases below each transition~\cite{13,14}.\\
The different Ce valence states in CeRuSn are intimately connected to the distances between a particular Ce ion and its Ru ligands. For smaller $d_{\rm{Ce-Ru}}$ distances, the overlap of the Ce 4{\it f\/} and Ru 4{\it d\/} electron wave-functions is larger and consequently the 4{\it f\/}--ligand hybridization is stronger. For $d_{\rm{Ce-Ru}}$ decreasing below a certain critical value, the Ce$^{3+}$ becomes gradually more admixed with the Ce$^{4+}$ state yielding an intermediate valence state. In the strong 4{\it f\/}--ligand hybridization limit, the 4{\it f\/}--electron becomes itinerant and the particular Ce ion becomes tetravalent.\\
The formation of new superstructures and coexistence of two or more Ce sites with different valence states in CeRuSn was predicted by ab initio band-structure calculations performed for CeRuSn by Matar {\it et al.}~\cite{11} and was corroborated also by results of $^{119}$Sn NMR and M{\"o}ssbauer spectroscopy measurements~\cite{15}.\\
A detailed neutron powder diffraction study~\cite{16} devoted to the investigation of the low temperature magnetic ordering discussed below also provides evidences that a certain volume fraction of the high-temperature 2{\it c}-superstructure phase (HTP) existed in the CeRuSn polycrystalline powder sample even at low temperatures with the low-temperature 3{\it c}-superstructure phase (LTP) so that the sample appeared "structurally" multiphase although remaining "chemically" single phase. This situation may arise e.\,g.\,, from the possibility that HTP in a certain volume fraction near to surface of powder particles survives during cooling as a metastable phase, whilst the bulk of the sample undergoes successively the transition from HTP to the intermediate-temperature 5{\it c}-superstructure phase (ITP). Generally, in polymorphic systems HTP may survive as a metastable phase while cooling through the equilibrium HTP $\leftrightarrow$ LTP transition temperature in a part of volume fraction or in the entire sample even in single crystals depending on thermal history~\cite{17,18} and only a proper heat treatment enables the formation of the entire stable LTP~\cite{19}. The large temperature hysteresis is presumably a consequence of high energy cost in order to change the atomic layer stacking as observed e.\,g.\,, in {\it RE}Ir$_2$Si$_2$ compounds~\cite{19} or due to displacement of the equilibrium atomic positions in CeRuSn by the structure transition between two polymorphs.\\
With the high-quality CeRuSn single-crystals being carefully prepared as concerns growth and heat treatment, detailed measurements of magnetization, electrical resistivity and thermal expansion in conjunction with X-ray single-crystal diffraction measurements at relevant temperatures clearly revealed that CeRuSn undergoes two distinct intrinsic first-order transitions, namely HTP $\leftrightarrow$ ITP and ITP $\leftrightarrow$ LTP, both of which exhibit a large temperature hysteresis~\cite{14}. By cooling, each transition yields a sudden contraction of the crystal along the {\it c}-axis (about 0.9~\% in total across the two transitions) as has been evidenced by thermal expansion and single-crystal X-ray diffraction measurements~\cite{13,14}.\\
After  closer inspection of Ce-Ru distances determined by the X-ray single crystal diffraction experiments~\cite{13,14}, one may conclude that both, HTP and LTP, contain one Ce$^{3+}$ ion per superstructure unit cell, whereas in ITP there exist two Ce$^{3+}$ ions, which are surrounded by Ru ligands at a long distance. The other Ce ions are expected being tetravalent and/or in an intermediate valence state, which is caused by a strong 4{\it f\/}--ligand hybridization because of the short Ce-Ru bonds.\\
An antiferromagnetic ordering in CeRuSn at low temperatures has been reported by Mydosh {\it et al.}~\cite{12} presenting results of measurements on polycrystalline samples. The magnetic ordering transition at $T_{\rm N} = 2.7$~K has been concluded from anomalies in the temperature dependence of the magnetization and the specific heat at this temperature. Recently published results of extensive neutron powder diffraction experiments~\cite{16} confirm that CeRuSn is antiferromagnetic below $T_{\rm N} = 2.8$~K with two incommensurate propagation vectors $\mathbf{q}_1 = (0,0, \approx 0.30)$ and $\mathbf{q}_2 = (0,0, \approx 0.43)$ with respect to the doubled crystal structure variant. It is argued that the reported neutron diffraction data can be best explained by a complicated distorted cycloidal magnetic structure with Ce moments of $\approx 0.7 \mu_{\rm B}$ being confined to the {\it a-c} plane~\cite{16}. However, the nature of the magnetic ground state is still far from being understood.\\
For revealing specific aspects of the low-temperature magnetism in this unusual material, namely the type of magnetic ordering, magnetocrystalline anisotropy and the role of the Ce ions residing in the crystallographically inequivalent sites and to shed more light on the nature of the magnetic ground state, relevant experiments performed on high quality single crystals are indispensable. To contribute to this process, we have grown several CeRuSn single crystals, characterized them thoroughly by X-ray diffraction and by EDX microprobe analysis and measured magnetization, specific heat and electrical resistivity at low temperatures in various magnetic fields applied along the principal crystallographic axes of the monoclinic structure.

\section{Experimental methodes}

The experiments discussed in this paper were done on the carefully characterized CeRuSn single crystals prepared by Czochralski method in a tri-arc furnace. Details on preparation and characterization of the crystals are available in Ref.~\cite{14}.\\
The magnetization $M$ was measured on a single crystal along the three main crystallographic axes of the monoclinic structure. Measurements in fields up to 7~T were realized utilizing a SQUID magnetometer (MPMS 7~T instrument of Quantum Design) in the temperature range between 1.8 and 20~K. The range of applied magnetic fields has been extended to 14~T using a PPMS 14~T instrument (Quantum Design) with the vibrating sample magnetometer option. The electrical resistivity  and the specific heat $C_{\rm p}$ were measured between 0.5 and 20~K using PPMS 14~T and PPMS 9~T, respectively. For the electrical resistivity measurements, the four-probe geometry was used with gold wires (25 $\mu$m in diameter) spot-welded to the samples. The dramatic structure transitions in CeRuSn when cooling from room temperature~\cite{13,14} are accompanied with formation of various cracks preventing proper determination of geometrical factor usually derived from bulk dimensions of bar shaped samples. Therefore, we display only the relative resistivity  $\rho/\rho_{\rm{300K}}$ data. For specific-heat measurements, the relaxation method implemented in PPMS instruments was employed. The experiments in magnetic fields have been done with field applied along each direction of the three principal crystallographic axes in order to collect important information on magnetic anisotropy.

\section{Results and discussion}

The temperature dependence of magnetic susceptibility $(M/H)$ below 10~K is displayed in Fig.~1 for all three main crystallographic directions. No difference has been found between the corresponding ZFC and FC dependencies, thus we show only the latter ones. In low magnetic fields, we have observed a maximum at $\sim 2.9$~K for the field applied along the $a$ and $c$--axis. This is a somewhat higher temperature than $T_{\rm N}$ suggested in literature~\cite{12} obtained from specific heat data (2.7~K). The temperature of specific heat maximum is usually considered to coincide with the temperature of magnetic ordering, in our case, the N\'{e}el temperature of the antiferromagnet CeRuSn. As shown by Fisher~\cite{21}, the correct value of $T_{\rm N}$ can be derived from susceptibility data being the temperature at which the product $(T \partial \chi /\partial T)$ has the maximum. In our case we have obtained values of 2.8~K and 2.7~K from analyzing the $c$- and $a$--axis data. \\

The $b$--axis susceptibility is rather weak and shows no maximum but only saturation below 2.9~K. This behavior remains intact in various magnetic field applied along the $b$-axis, which is general behavior in the hard magnetization direction of an anisotropic antiferromagnet.
The 2.9~K maximum observed in the $M/H$ vs. $T$ plots in low fields applied along the $a$ and $c$--axis, shifts to lower temperatures with increasing magnetic field and, for fields higher than a certain critical value, the maximum vanishes. Such behavior is typical for antiferromagnets with the susceptibility showing a maximum for $H \rightarrow 0$ in the vicinity of the N\'{e}el temperature ($T_{\rm N}$).
The magnetocrystalline anisotropy within the $a-c$ plane is clearly manifested also by the fact that the temperature of the susceptibility maximum develops differently depending on the direction of the magnetic field. In fields along the $c$--axis (lower panel Fig.~1) the shift of the maximum towards lower temperatures becomes more rapid above  $\mu_0H > 0.5$~T in comparison to the $a$--axis.\\

As one can see from the magnetization isotherms measured at 1.8~K shown in Fig.~2, CeRuSn undergoes a metamagnetic transition in the magnetic field applied along the $a$ and $c$--direction with critical fields $\mu_0H_{\rm c} \sim 1.5$~T and 0.8~T, respectively, showing no hysteresis. The $\mu_0H_{\rm c}$ value represents the magnetic field at which the $\partial M/\partial H$ has a maximum. The metamagnetic transition is also typical for antiferromagnets and signals the moment where the original antiferromagnetic structure becomes modified or completely destroyed by (partially) aligning the magnetic moments along the field direction. The metamagnetic transition is usually observed in the field direction for which the temperature dependence of the magnetic susceptibility shows a maximum as in our case. The magnetization of CeRuSn in the fields above the metamagnetic transition shows a strong tendency to saturation. It saturates well for field $\geq 4$~T in the $c$--axis direction. In the magnetic field applied along the $b$--direction, for which the temperature dependence of magnetic susceptibility shows no maximum, the magnetization shows paramagnetic response remaining very small even at 14~T. The magnetic moment determined at 14~T reaches the values 0.51, 0.08 and 0.71 $\mu_{\rm B}/f.u.$ for the $a$, $b$ and $c$--axis, respectively, confirming strong anisotropy in the ordered state with the $c$--axis ($b$--axis) as the easy (hard) magnetization direction. \\
The ordered magnetic moment of a free Ce$^{3+}$ ion equals to $g_{\rm J}\cdot J = 2.16 \mu_{\rm B}$. The saturated $c$--axis value of $0.71 \mu_{\rm B}/f.u.$ approximately corresponds to one third of 2.16, which is consistent with the scenario of the CeRuSn low-temperature polymorphic phase containing only one Ce$^{3+}$ ion of the totally three Ce ions in the unit cell. The remaining two cerium ions in CeRuSn are considered carrying no stable magnetic moment since they are either tetravalent or in an Ce$^{3+}$--Ce$^{4+}$ intermediate valence state due to strong 4{\it f\/}(Ce)-4{\it d\/}(Ru) hybridization, which is induced by a very short $d_{\rm{Ce-Ru}} \sim 2.3$~\AA. In contrary, the Ce$^{3+}$ ion has the Ru ligands at a long distance ($\sim 2.9$~\AA). The magnetic moments of the nearest neighbor Ce$^{3+}$ ions along the $c$--axis are coupled along the $c$--axis by a rather weak and complex long-range exchange interaction mediated by the conduction electrons (RKKY-type), which may be assisted by a multiple Ce-Ru hybridization in the Ce$^{3+}$--Ru--Ce--Ru--Ce--Ru--Ce$^{3+}$ chain winding along the $c$--axis (see crystal structure figures in Ref.~\cite{14}).
The lowest $\mu_0H_{\rm c}$ value found along the $c$--axis, as well as the higher well saturated magnetization value, may arise from an antiferromagnetic structure built of antiparallel coupled magnetic moments aligned along or close to the $c$--axis. The $c$--axis metamagnetic transition would then consist of a spin flip process in a field sufficient to break the antiferromagnetic coupling and aligning the moments along the field direction.\\
Also the metamagnetic transition in the $a$--axis is closely related to a destruction of the ground-state antiferromagnetic ordering. Nevertheless a field of 14~T is insufficient to align the magnetic moments along the field direction. This reflects the magnetic anisotropy within the $a-c$ plane. A new non-collinear magnetic structure is formed stabilized by the $a$--axis magnetic field and yielding a moment of $0.51 \mu_{\rm B}/f.u.$ at 14~T.
The weak response of the magnetization to the magnetic field applied along the $b$--axis, contrary to the metamagnetic transition for the $a$ and $c$--axis, may be conceived within a scenario of an antiferromagnetic structure having the Ce magnetic moments fixed perpendicular to this axis, i.\,e.\,, within the $a-c$ plane. This easy-plane anisotropy is so strong that the moments cannot be deflected from the $a-c$ plane toward the $b$--axis by a field of 14~T. The features of the magnetic structure of CeRuSn deduced from our magnetization data are quite in agreement with the recently published results of neutron powder diffraction experiments~\cite{16}. \\
To reveal all the details of the magnetic structures in CeRuSn (ground-state and field-induced phases), combined polarized neutron and resonant X-ray single crystal magnetic diffraction in magnetic fields supported by results of relevant XMCD experiments on well-defined single crystals are strongly desired.\\
The specific heat of CeRuSn shown in Fig.~3 exhibits a sharp anomaly having a maximum at 2.8~K, which can be attributed to the transition from a paramagnetic phase to an antiferromagnetic ordering. Integration of the $C_{\rm p}/T$ vs. $T$ dependence over temperature yields the entropy, which in overall reaches a value of $0.32 R\ln2$ at 4~K. At this temperature the entropy is almost entirely of magnetic origin. This means that the magnetic entropy in CeRuSn at 4~K is close to $1/3 R\ln2$, which also corroborates the scenario based on only one third of cerium atoms bearing a magnetic moment equivalent to the moment of a free Ce$^{3+}$ ion. These Ce magnetic moments order antiferromagnetically below $T_{\rm N}$. Application of an external magnetic field in the $a$ and $c$--direction firstly shifts the specific heat anomaly to lower temperatures as expected for an antiferromagnet similarly to the evolution of the susceptibility maximum presented before. In fields higher than the critical field of the metamagnetic transition, the anomaly smears out and moves to higher temperatures, which we understand as arising due to a gradual aligning of the magnetic moments along the field direction. In contrary, the specific-heat anomaly is hardly influenced by a magnetic field of 9~T applied along the $b$--axis. This is consistent with the "paramagnetic" response in magnetization confirming the $b$--axis as the hard magnetization axis.\\

The temperature dependence of the electrical resistivity for current applied along the $a$--axis is depicted in Fig.~4. The antiferromagnetic phase transition of CeRuSn is accompanied by a sudden drop of resistivity below 3~K as a result of the reduced spin-dependent scattering due to the transition from an incoherent paramagnetic state to a coherent antiferromagnetic structure~\cite{22}. A similar temperature dependence of the resistivity in zero magnetic field was observed for the direction of electrical current along the other two principal crystallographic axes (not shown). The magnetic field applied in the current direction along the $a$ (see Fig.~4) or $c$--axis (not shown) leads to a gradual smearing out of the $T_{\rm N}$ related anomaly.\\
We have also measured the longitudinal magnetoresistance to see especially the response of electrical transport to the field induced metamagnetic transitions. Generally, the metamagnetic transitions of antiferromagnets are accompanied by magnetoresistance anomalies~\cite{22,23,24}. In Fig.~5, the longitudinal magnetoresistance curves measured for the current along the $a$--axis at 1.8 and 0.5 K are displayed. Two anomalies occur in the neighborhood of the critical field $H_{\rm c}$ of the metamagnetic transition observed on the magnetization curve (see Fig.~2). On the 1.8~K curve, a peak spreading from 0~T to $H_{\rm c}$ is followed by a large negative response in fields above $H_{\rm c}$ (amounting - 25\% above 6~T). This latter features correlate with the magnetization above the metamagnetic transition. The peak in fields just below $H_{\rm c}$ is usually interpreted in terms of an additional scattering of conduction electrons on spin-flip fluctuations from the coherent ground state antiferromagnetic structure towards the field aligned metamagnetic state. This can be seen in terms of the simple N\'{e}el model with two magnetic sublattices, where the dominating effect is the loss of periodicity due to the reversal of moments (spin flips) residing in the sublattice with antiparallel orientation~\cite{22}. Similar positive magnetoresistance for $H < H_{\rm c}$ is expected also for band antiferromagnets~\cite{25}.\\
When $H_{\rm c}$ is reached, the spin-flip fluctuations cease, the positive magnetoresistance vanishes and the metamagnetic transition (seen on the magnetization curve in Fig.~2) has a large negative magnetoresistance response (Fig.~5). Qualitatively similar longitudinal magnetoresistance behavior is observed also for the $c$--axis current (not shown). These results are analogous to magnetoresistance behavior of many antiferromagnets in the neighborhood of a metamagnetic transition and are generally understood as follows: the antiferromagnetic spin structure usually has a different periodicity than the underlying crystal lattice. Consequently the exchange interaction between the conduction electrons and magnetic ions introduces new boundaries in the Brillouin zone forming a gap over a portion of the Fermi surface. The metamagnetic transition changes the antiferromagnetic structure to a state with magnetic moments aligned along the field direction, which has a magnetic periodicity coinciding with the crystallographic one. Consequently the Fermi surface is reconstructed (the gap is removed) causing a considerable resistivity decrease~\cite{23,24,25}.\\
A much smaller and narrower peak on the magnetoresistance curve measured at 0.5~K reflects a reduced influence of thermal fluctuations in generating the spin-flip fluctuations at lower temperatures. The magnetoresistance drop due to the metamagnetic transition becomes very sharp in $T = 0.5$~K. Considering the close correlation between the magnetoresistance and magnetization in fields above $H_{\rm c}$~\cite{23,24,25}, one may expect also a sharp step of the magnetization at the metamagnetic transition at 0.5~K. At the moment, however, we cannot confirm this conclusion having no magnetization measurements at temperatures below 1.8~K available in our laboratory.\\

\section{Conclusions}
We have studied low-temperature magnetization, specific heat and electrical resistivity of CeRuSn single crystals as functions of temperature and the magnetic field applied along the principal crystallographic directions in order to shed more light on the complicated magnetism of a material containing Ce ions in more than one valence state. Our data confirm anisotropic antiferromagnetic ordering ($T_{\rm N} = 2.8$~K) of Ce magnetic moments locked within the $a-c$ plane leaving the $b$--axis as the hard magnetization direction. The antiferromagnetic structure is destroyed by a metamagnetic transition in the magnetic field applied along the $a$ or $c$--axis.\\
The saturated easy-axis ($c$--axis) magnetization, as well as the magnetic entropy above $T_{\rm N}$, corresponds to a scenario with one third of Ce ions in CeRuSn carrying the magnetic moment equivalent to a free Ce$^{3+}$ ion, whereas the other two thirds of Ce ions are in a nonmagnetic mixed valence or tetravalent state. This is consistent with the expected low temperature CeRuSn crystal structure (a superstructure consisting of three unit cells of the CeCoAl-type piled up along the $c$--axis also called the $3c$ superstructure), in which the Ce$^{3+}$ ions are characteristic by large distances from the nearest Ru ligands, whilst the Ce-Ru distances of the other two thirds of Ce ions are much shorter. The latter fact leads to a strong 4{\it f\/}--ligand hybridization and consequently tetravalent and/or mixed valence state of the corresponding Ce ions. To our best knowledge, CeRuSn is a unique case among Ce intermetallics, in which a part of Ce ions being in Ce$^{3+}$ state coexists in the crystal lattice with mixed valence or tetravalent Ce ions with the Ce valence being controlled by the Ce-ligand hybridization.\\
The antiferromagnetic structure is apparently very complex and arises from an intricate hierarchy of exchange interactions mediated by conduction electrons together with a multiple Ce-Ru hybridization in the Ce$^{3+}$--Ru--Ce--Ru--Ce--Ru--Ce$^{3+}$ chain winding along the $c$--axis interplaying with the crystal field interaction.

\section*{Acknowledgements}
This work was supported by the Czech Science Foundation (grant GACR 202/09/1027) and the Charles University (grant GAUK 440811). Experiments were performed in MLTL (http://mltl.eu/), which is supported within the program of Czech Research Infrastructures (project no. LM2011025).\\

\begin{figure}[ht]
\centerline{\includegraphics[width=0.8\textwidth]{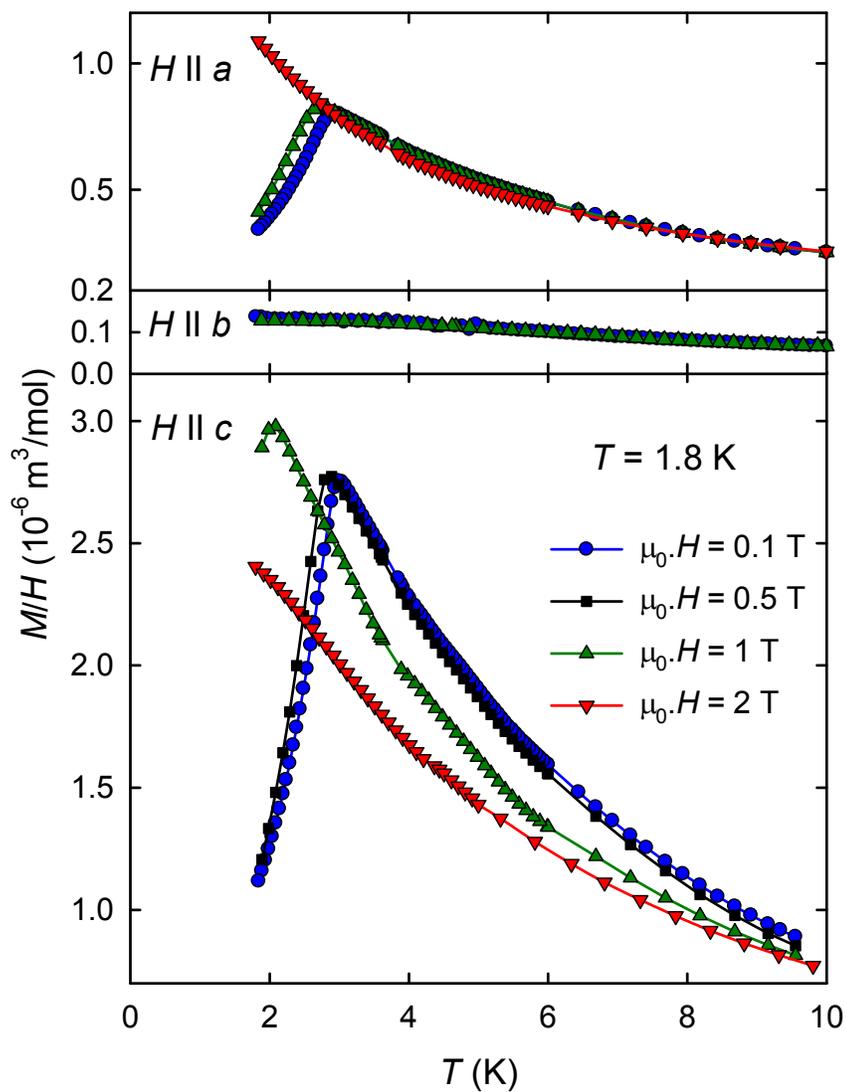}}
\caption{\label{Fig1} Temperature dependence of magnetic susceptibility $(M/H)$ at temperatures below 10~K measured in various magnetic fields applied along the principal crystallographic directions.}
\end{figure}

\begin{figure}[ht]
\centerline{\includegraphics[width=0.8\textwidth]{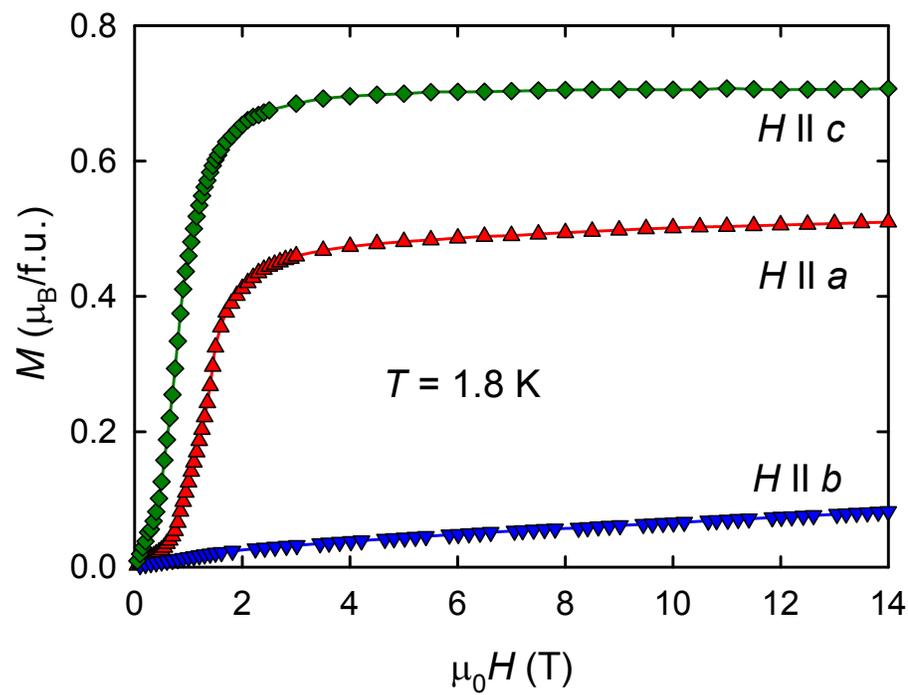}}
\caption{\label{Fig2} Magnetization isotherms at 1.8~K measured along the principal directions.}
\end{figure}

\begin{figure}[ht]
\centerline{\includegraphics[width=0.8\textwidth]{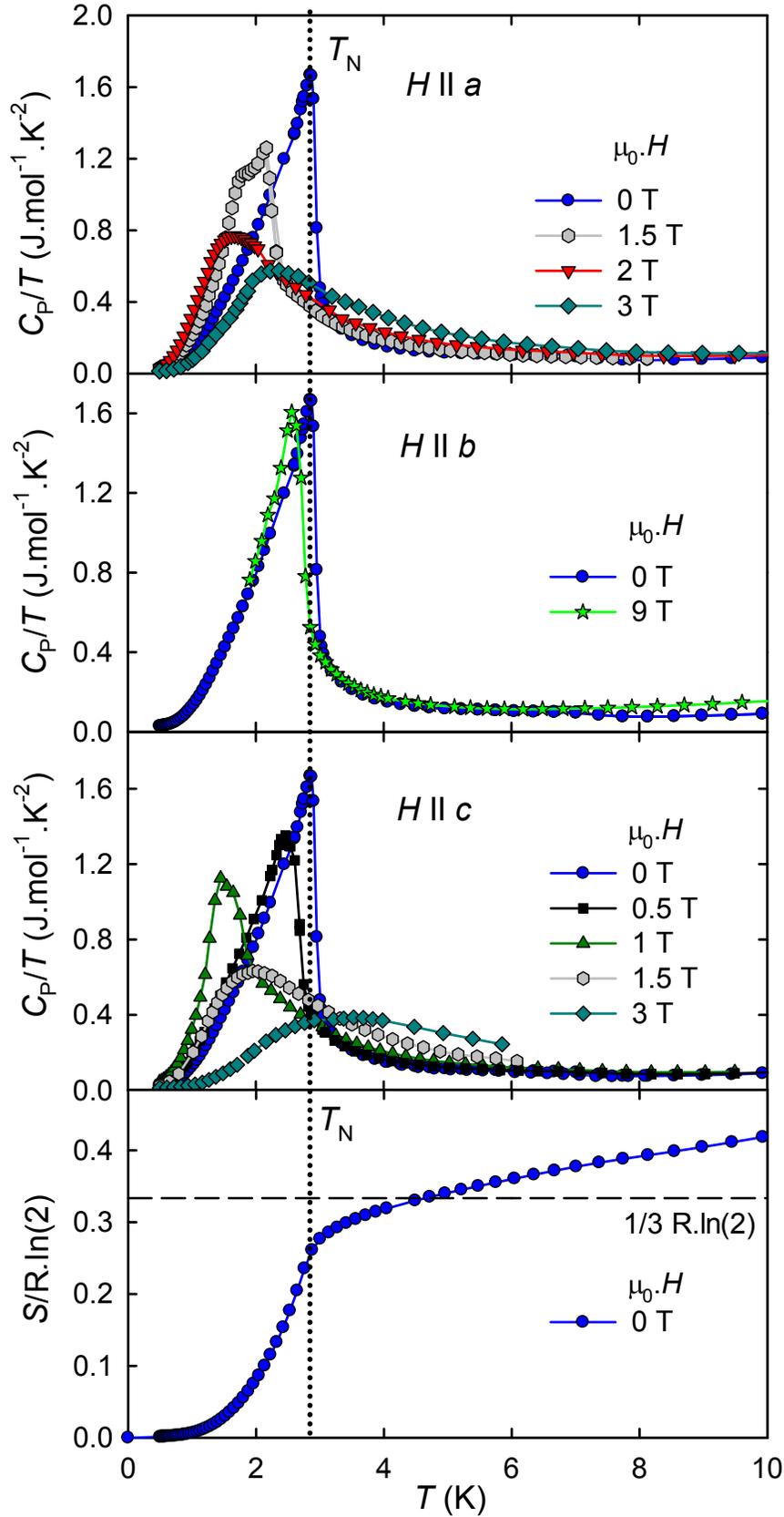}}
\caption{\label{Fig3} Temperature dependence of specific heat ($C_{\rm p}/T$ vs. $T$ plot) of CeRuSn exposed to various fields applied along the $a$, $b$, and $c$--axis. The bottom panel displays the entropy in zero magnetic field. The vertical dotted line marks the N\'{e}el temperature.}
\end{figure}

\begin{figure}[ht]
\centerline{\includegraphics[width=0.8\textwidth]{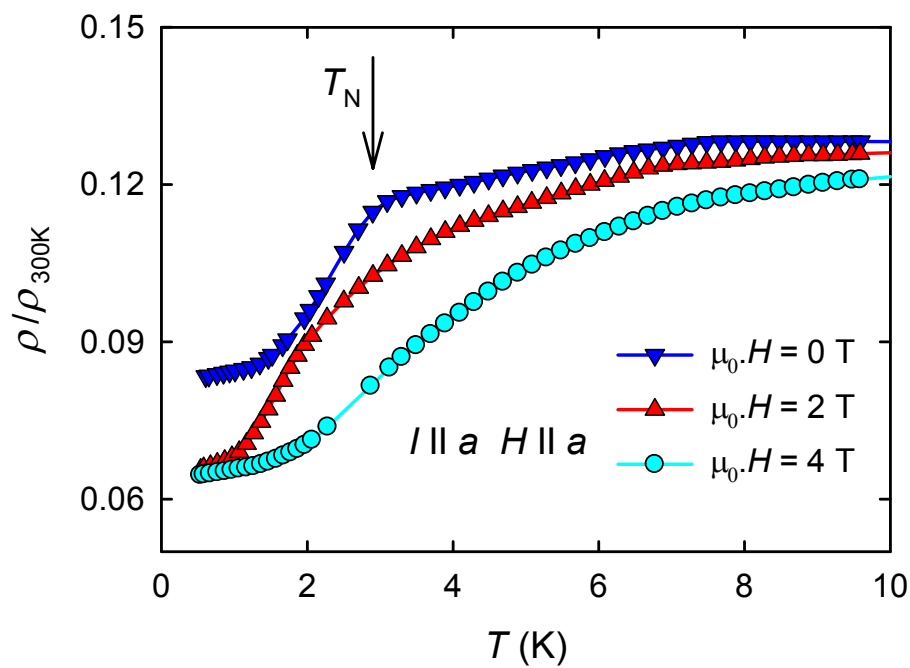}}
\caption{\label{Fig4} Low-temperature scans of the relative resistivity for a current along the $a$--axis measured in the field of 0, 2 and 4~T applied along the current direction. $T_{\rm N}$ marks the N\'{e}el temperature determined from specific heat data.}
\end{figure}

\begin{figure}[ht]
\centerline{\includegraphics[width=0.8\textwidth]{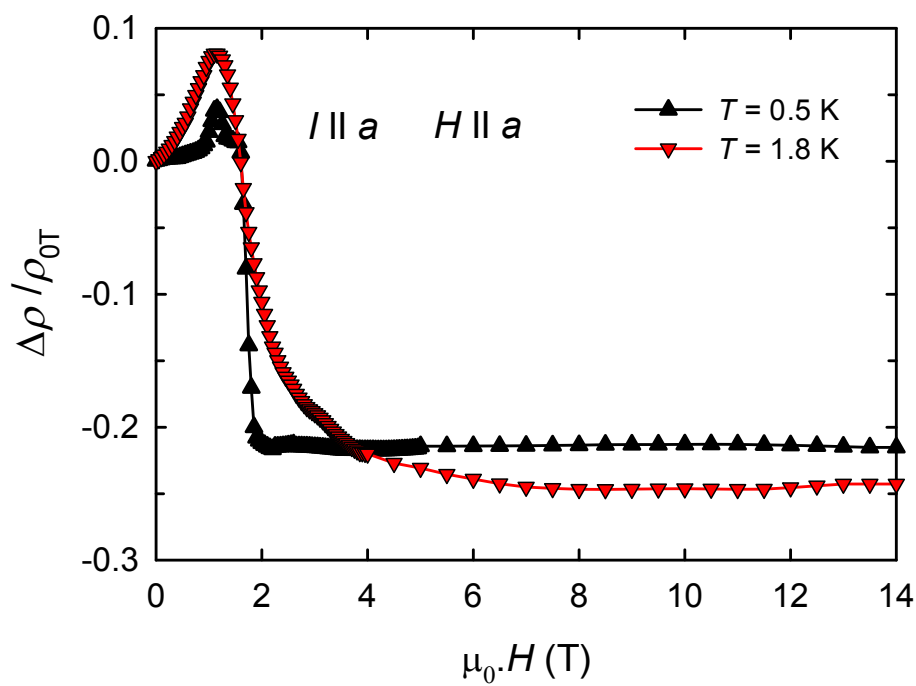}}
\caption{\label{Fig5} Longitudinal magnetoresistance of CeRuSn for current along the $a$--axis measured at 0.5 and 1.8~K. }
\end{figure}

\end{document}